\documentclass[usegraphicx,usenatbib]{mn2e}
\usepackage{times}
\def\mpc{h^{-1} {\rm{Mpc}}}

\def\apj {ApJ}
\def\apjl {ApJL}
\def\apjs {ApJS}
\def\aj {AJ}
\def\aap {A\&A}
\def\mnras {MNRAS}
\begin{document}
\title[Red galaxy sequence]
{The environmental dependence of the red galaxy sequence}
\author[H.J. Mart\'{\i}nez, V. Coenda \& H. Muriel]
{H\'ector J. Mart\'\i nez,\thanks{E-mail: julian@oac.uncor.edu} 
Valeria Coenda \& Hern\'an Muriel\\
Instituto de Astronom\'{\i}a Te\'orica y Experimental (IATE), 
CONICET$-$Observatorio Astron\'omico, Universidad Nacional de 
C\'ordoba,\\ Laprida 854, X5000BGR, C\'ordoba, Argentina}
\date{\today}
\pagerange{\pageref{firstpage}--\pageref{lastpage}} 
\maketitle
\label{firstpage}
%%%%%%%%%%%%%%%%%%%%%%%%%%%%%%%%%%%%%%%%%%%%%%%%%%%%%%%%%%%%%%%%%%%%%%%%%%%%%%%
%%%%%%%%%%%%%%%%%%%%%%%%%%%%%%%%%%%%%%%%%%%%%%%%%%%%%%%%%%%%%%%%%%%%%%%%%%%%%%%
\begin{abstract}
The dependence of the sequence of red galaxies (RS) with the environment is investigated using
volume limited samples of field, group, and cluster galaxies drawn from the Sloan 
Digital Sky Survey (SDSS).
Our work focuses in studying the mean colour ($\mu_R$) and the scatter ($\sigma_R$) of 
the RS as a function of $^{0.1}r-$band absolute magnitude in different environments characterised either
by the mass of the system in which the galaxies are located or by the distance to
the system's centre. These results are compared with the RS of field galaxies.
The same analysis is carried out using subsamples of red galaxies classified as early types
according to their concentration parameter. 
For a given luminosity, $\mu_R$ of field galaxies is bluer and $\sigma_R$ is larger
than their group and cluster counterparts irrespective of mass and position within the systems.
Among systems of galaxies, high mass groups and clusters have the reddest $\mu_R$ and the smallest
$\sigma_R$. 
These differences almost disappear when red early type galaxies alone are considered.
Galaxies in the core and in the outskirts of groups have similar $\mu_R$, whereas galaxies in 
clusters show a strong dependence on cluster centric distance, being the 
inner galaxies the reddest objects. Red early type galaxies in the outskirts of clusters
have $\sigma_R$ values as large as field galaxies', while galaxies in the inner regions of
clusters have lower values, comparable to those of group galaxies.  
We find that bright red early type galaxies have reached nearly the same evolutionary stage in
all environments. Our results suggest that, although effective in drifting galaxies
of intermediate luminosities towards redder colours, the cluster environment is not 
necessary to populate the RS. We propose a scenario in which the RS in massive systems is populated by
two different star formation history galaxies: red early type galaxies that
formed the bulk of their stars during the early stages of massive halo assembly, and
red galaxies that passed most of their lives inhabiting poor groups or the field and 
fell into massive systems at lower redshifts.
\end{abstract}
%%%%%%%%%%%%%%%%%%%%%%%%%%%%%%%%%%%%%%%%%%%%%%%%%%%%%%%%%%%%%%%%%%%%%%%%%%%%%%%
%%%%%%%%%%%%%%%%%%%%%%%%%%%%%%%%%%%%%%%%%%%%%%%%%%%%%%%%%%%%%%%%%%%%%%%%%%%%%%%
\section{Introduction} 

Several years of research have demonstrated that galaxy properties strongly depend
on mass, time and environment. Galaxy properties are related in different ways
with these three fundamental parameters. Colour and luminosity have been proposed as two
of the properties that predict best the environment \citep{blanton05,hm2,mcm08}.
Nevertheless, for bulge-dominated galaxies, these two properties show a very robust correlation: the RS,
also known as colour magnitude relation (CMR, \citealt{vs77}).
\citet{faber73} and \citet{bower92} found a high degree of
universality for this relation comparing early type galaxies in clusters, groups and the
field. The colour of a galaxy correlates with the ratio of the present star formation rate
(SFR) to the overall mass of stars in the galaxy. Since the present day's SFR is several
orders of magnitude smaller than in the past, nearly half of the stellar mass of the
universe is in galaxies that populate the RS \citep{bell04}.
The RS is primarily the result of a relation between mass and metallicity
\citep{faber73}. Bright galaxies are redder because they are more metal enriched,
producing the well known slope of the RS. Age variations may also play a role \citep{faber73}
being the reddest galaxies the older population.

There is a lot of discussion going on about how the RS is populated. It has been suggested 
that galaxies move from the blue population (blue cloud) to the RS due to the truncation of the star
formation in $L<L^{\ast}$ galaxies \citep{bell07}. One of the proposed mechanisms to
produce this effect is the merging of galaxies in the framework of the hierarchical galaxy
formation. For the most massive galaxies, the proposed model involves mergers of galaxies
that already are in the RS (dry mergers, \citealt{vandokkum05}). Several physical
mechanisms associated with the environment can also play an important role in quenching the
star formation, like for instance the hostile cluster environment that can suppress the star
formation. \citet{tanaka05} suggest that the bright end  of the cluster RS is
already in place at $z\sim1$ while this relation for field galaxies has been built all the way
down to the present day. 
Using a semi-analytic model, \citet{delucia07} also suggest that the RS seems
to be in place at intermediate redshifts, nevertheless, they found a significant deficit of
faint red galaxies in clusters at $z\sim0.8$. suggesting that the RS population of high redshift
clusters does not contain all the progenitors of nearby RS cluster galaxies. 
These authors found differences in the formation times of the bulk of stars between cluster and field 
galaxies. Similar results were found by \citet{kav09}. Nevertheless, 
\citet{menci08}, also using a semi-analytic model, found 
similar mass growth history of field and cluster galaxies at $z\sim 1.2$, although, a larger 
dispersion of ages for RS field galaxies was reported. Based on observational data, 
\citet{stott07} and \citet{stott09} investigated the evolution of the RS in clusters
from $z\sim 1$ to the present and found that the rest-frame slope of the RS relation
evolves with redshift, this is attributed to the build up of the RS over time. 
Analysing a sample of 127 rich galaxy clusters \citet{lu09} found no 
strong evolution of the red-sequence's dwarf-to-giant ratio in the range $0.2 \leq z \leq 0.4$ but 
they found a significant increase of the red-sequence dwarf galaxies
from $z\sim 0.2$ to $z\sim0$. \citet{krick08} also found a deficit of faint galaxies on 
the red sequence in clusters at $z\sim1$, suggesting that, in clusters, the most massive galaxies 
have evolved faster than lesser massive galaxies.
On the contrary, \citet{andreon05} suggested that the abundance of red galaxies, 
relative to bright ones, is constant over the redshift range $0<z<1.3$.

Although the most massive galaxies that populate the bright region of the RS reside in high
mass clusters, galaxies in different environments seem to show similar RSs. \citet{hogg04}
analysed a sample of 55158 galaxies taken from the SDSS and found that the
colour-magnitude relation is independent of the of environmental overdensity.
Analysing galaxies at $z\sim0.7$, \citet{cassata07} found that the slope of the RS
is independent of the local density, Similarly, \citet{stott09}
observed no strong correlation between this slope and the cluster environment at a given
redshift. \citet{mac05} analysed the RS of three Abell clusters of 
different masses and found that these clusters have similar RSs and narrow limits of intrinsic 
colour scatter and slope. \citet{lcruz04} analysed a sample of 57 
X-ray-detected Abell clusters of different richness and cluster types and found that they 
present similar RSs.
On the other hand,  \citet{haines06} and  \citet{mei09}
found evidences of a relation between the RS and the environment in clusters of
galaxies. Analysing the core of the Shapley supercluster, \citet{haines06}
found that the RS is 0.015mag redder in highest-density regions. They
interpret this result in terms of ages, suggesting a population being 500-Myr older in the
cluster cores. 
\citet{pim02} and \citet{pim06} analysed a sample of 11 low redshift X-ray luminous 
clusters and found dependencies of the RS properties as a function of the projected cluster-centric
distance. For 12 intermediate-redshift X-ray 
clusters, \citet{wake05} found similar results, also interpreted in terms of age gradients.
\citet{mei09} also found evidences of age differences between galaxies
in the core and outskirts of clusters at high redshift. 
\citet{gilbank08} used composite samples of published cluster and field galaxies photometry 
and found that the dwarf-to-giant ratio in clusters is higher than that of the field at all 
redshifts, suggesting that the faint end of the RS was established in clusters. 
\citet{koyama07} analysed the deficit of faint red galaxies in clusters and found that this 
effect depends on cluster richness or mass, in the sense that poorer systems show stronger deficits.

The scatter around the RS can be used to constrain the way that galaxies reach the RS. In the
merger scenario, where  the most massive galaxies in the RS are the result of the merger of 
galaxies already in the RS \citep{vandokkum05}, a significant scatter is expected 
\citep{bower92}. Nevertheless, \citet{bernardi07} and \citet{skelton09} argued that
a small scatter is not against the dry merger scenario.   
The intrinsic scatter could also be the result of a spread in the age of the stellar population, 
as a result of differences in the time of the formation of the bulk of the stars at high 
redshift or the quenching of the star formation in the blue cloud at different epochs. 
\citet{mei09} systematically studied the RS scatter in high-redshift clusters as a function 
of different parameters. They found that the RS scatter for 
bright elliptical galaxies ($M_B<-21)$ increases with distance from the cluster centre, 
on the other hand, all the elliptical galaxies of fainter magnitudes show similarly 
larger scatter. These results are interpreted in terms of differences of ages.

%%%%%%%%%%%%%%%%%%%%%%%%%%%%%%%%%%%%%%%%%%%%%%%%%%%%%%%%%%%%%%%%%%%%%%%%%%%%%%%
%%%%%%%%%%%%%%%%%%%%%%%%%%%%%%%%%%%%%%%%%%%%%%%%%%%%%%%%%%%%%%%%%%%%%%%%%%%%%%%
\begin{figure}
\includegraphics[width=90mm]{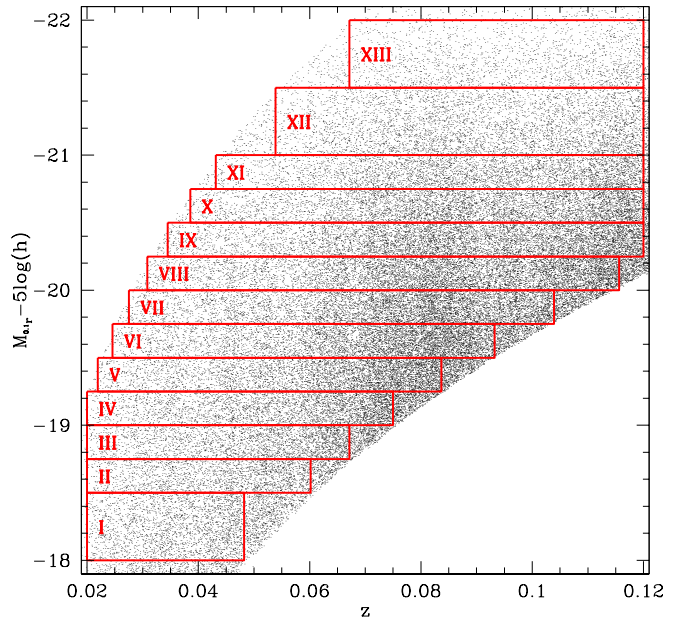}
\caption{Absolute $^{0.1}r-$band magnitude vs. redshift for galaxies in groups
in the SDSS DR7 used in this paper. Only galaxies with apparent magnitudes $14.5\le r\le17.77$
are shown. Each red rectangle bounds the area where the galaxy sample is volume limited
for galaxies within the corresponding luminosity bin (see text for details).}
\label{fig1}
\end{figure}
%%%%%%%%%%%%%%%%%%%%%%%%%%%%%%%%%%%%%%%%%%%%%%%%%%%%%%%%%%%%%%%%%%%%%%%%%%%%%%%
%%%%%%%%%%%%%%%%%%%%%%%%%%%%%%%%%%%%%%%%%%%%%%%%%%%%%%%%%%%%%%%%%%%%%%%%%%%%%%%
%%%%%%%%%%%%%%%%%%%%%%%%%%%%%%%%%%%%%%%%%%%%%%%%%%%%%%%%%%%%%%%%%%%%%%%%%%%%%%%
%%%%%%%%%%%%%%%%%%%%%%%%%%%%%%%%%%%%%%%%%%%%%%%%%%%%%%%%%%%%%%%%%%%%%%%%%%%%%%%
\begin{figure}
\includegraphics[width=90mm]{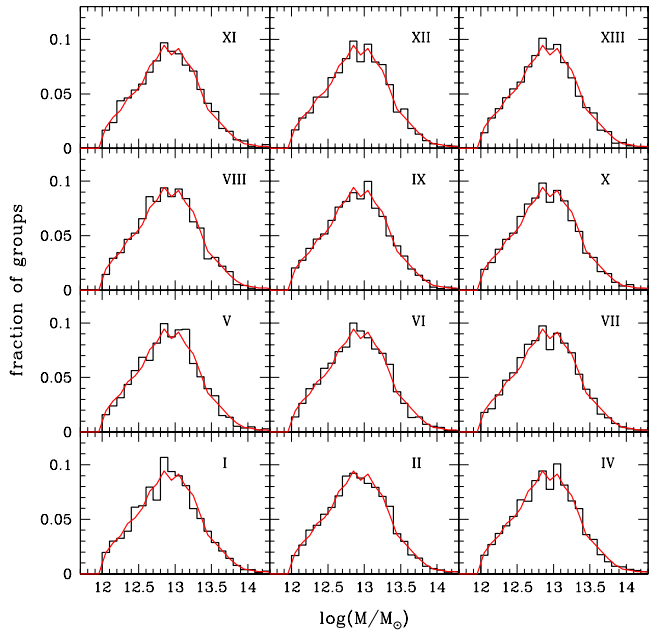}
\caption{The mass distribution of groups contributing galaxies to the
luminosity bins denoted by capital Roman numbers, that correspond to 
the rectangles in Figure \ref{fig1}. Groups in each panel have been
selected using a Monte-Carlo algorithm in order to match the mass distribution
shown in all panels as red continuous line that corresponds to the III luminosity bin 
(see text for details).}
\label{fig2}
\end{figure}
%%%%%%%%%%%%%%%%%%%%%%%%%%%%%%%%%%%%%%%%%%%%%%%%%%%%%%%%%%%%%%%%%%%%%%%%%%%%%%%
%%%%%%%%%%%%%%%%%%%%%%%%%%%%%%%%%%%%%%%%%%%%%%%%%%%%%%%%%%%%%%%%%%%%%%%%%%%%%%%
\begin{figure}
\includegraphics[width=90mm]{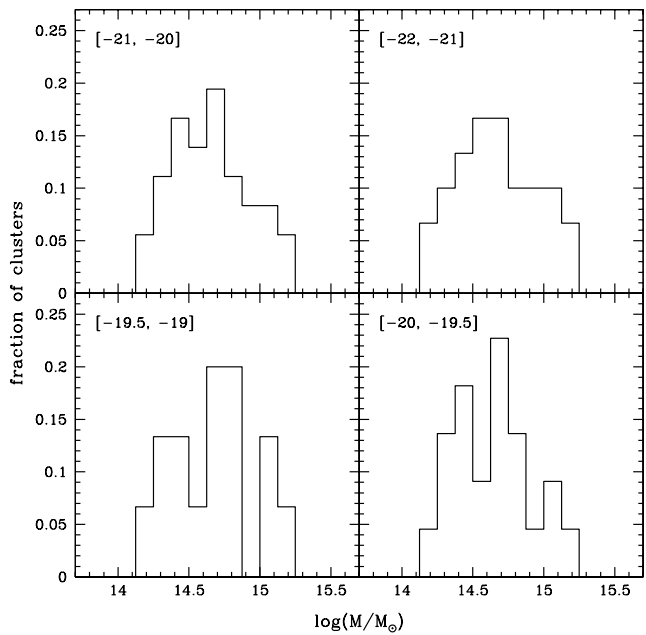}
\caption{The mass distribution of X-ray clusters that contribute galaxies to
the luminosity bins indicated in each panel.}
\label{fig3}
\end{figure}
%%%%%%%%%%%%%%%%%%%%%%%%%%%%%%%%%%%%%%%%%%%%%%%%%%%%%%%%%%%%%%%%%%%%%%%%%%%%%%%
%%%%%%%%%%%%%%%%%%%%%%%%%%%%%%%%%%%%%%%%%%%%%%%%%%%%%%%%%%%%%%%%%%%%%%%%%%%%%%%

In this paper, we probe the environmental dependence of the RS by analysing the
mean colour and scatter of the RS as a function of galaxy absolute magnitude in samples
of galaxies in the field, in groups and in X-ray clusters. The aim of this work is to determine
a precise RS for galaxies in different environments using a uniform sample of galaxies, and to 
shed light on the understanding of its origin.
In groups and clusters we analyse how the RS depends on system mass and the 
projected distance to the centres of the systems. 
This paper is organised as follows: section 2 describes the samples of galaxies used;
the dependence of the RS on mass and system-centric distance is analysed in section 3;
in section 4 we summarise and discuss our results.
%%%%%%%%%%%%%%%%%%%%%%%%%%%%%%%%%%%%%%%%%%%%%%%%%%%%%%%%%%%%%%%%%%%%%%%%%%%%%%%
%%%%%%%%%%%%%%%%%%%%%%%%%%%%%%%%%%%%%%%%%%%%%%%%%%%%%%%%%%%%%%%%%%%%%%%%%%%%%%%
\section{The samples of galaxies}
To probe the environmental dependence of the RS, we use samples of field, group and
cluster galaxies drawn from the Main Galaxy Sample (MGS; \citealt{mgs}) of the
last public release of the SDSS (DR7; \citealt{dr7}).
In some of the analyses below, we classify galaxies into early and late types
according to their concentration parameter \citep{st01}, thus we have restricted our
galaxy samples to be  $z \le 0.12$ as a compromise between the number of objects and avoiding 
non-negligible seeing  effects in the values of the radii enclosing 50 and 90 per cent of Petrosian flux.

Groups of galaxies in the DR7 MGS used in this work were identified by Dr. Ariel Zandivarez 
(private communication) 
using a standard friend-of-friend (\textit{fof}) technique as in \citet{mz05}. Given the known 
sampling problems for bright galaxies, galaxies with apparent magnitude 
$r<14.5$ were excluded before running the finder algorithm, i.e., the group identification
was carried out over all MGS galaxies with $14.5\le r\le 17.77$. 
We refer the reader interested in the details to the \citet{mz05} paper.
The following values were used in the group identification: 
a transverse linking length  corresponding to an overdensity $\delta\rho/\rho=200$,
a line-of-sight linking length $V_0=200 {\rm km~s^{-1}}$ (both of them scale with
a factor that compensates for the decline of the selection function with distance
that assumes the galaxy luminosity function found by \citet{blanton03b}), the same cosmology used
in this paper, and a fiducial distance of $D_f=10\mpc$. 
The identification resulted in 11028 groups with at least 4 galaxy members, adding up to a total of 
78056 galaxies in the redshift range of our interest $0.02\le z \le 0.12$. This sample has median virial mass 
$\cal{M}$$=1.5\times10^{13}h^{-1}\cal{M}$$_{\odot}$, median virial radius $R_{\rm vir}=0.8h^{-1}\rm{Mpc}$
and median redshift $z=0.08$. 
According to the analyses of group identification accuracy performed over mock catalogues
by \citet{mz02}, less than $\sim 25\%$ of groups identified with the parameters above should be spurious. 
It is worth mentioning that we have not excluded groups from the sample on the basis of matching them with
known clusters of any kind. 

Galaxies in X-ray selected clusters used in this paper are a subsample of the C-P04-I
sample of \citet{coenda09} who selected clusters from the ROSAT-SDSS Galaxy Cluster
Survey of \citet{popesso07}.
The C-P04-I sample comprises 42 regular galaxy clusters in the redshift range $0.05<z<0.12$.
Galaxies in these clusters were identified using the MGS of the Fifth Data
Release (DR5) of SDSS \citep{dr5}.
In order to compute cluster properties, \citet{coenda09} identified cluster members in two steps.
First, they used the \textit{fof} algorithm with the linking parameters and modifications
introduced by \citet{Diaz:2005}. Then they performed an eyeball examination of the
structures detected by \textit{fof}. From the redshift distribution of galaxies, 
they determined the line-of-sight extension of each cluster and therefore, the clusters members. 
Through visual inspection, \citet{coenda09} exclude systems that have two or more close substructures 
of similar size in the plane of the sky and/or in the redshift distribution. 
Finally, the authors computed cluster physical properties such as the line-of-sight velocity 
dispersion $\sigma$, virial radius and mass. They found a median virial mass of 
$\cal{M}$$_{vir}=4\times10^{14}$ $h^{-1}\cal{M}$$_{\odot}$ and a median virial radius of $R_{vir}=1.6$$\mpc$.

We consider as 'field galaxies' to all DR7 MGS galaxies that were
not identified as belonging to \textit{fof} groups nor to \citet{coenda09} clusters.
Thus, our field galaxy sample should be contaminated by galaxies in systems not
identified by the group finder algorithm. While interpreting the outcomes of our
analyses this fact should be kept in mind.

Galaxy magnitudes used throughout this paper have been corrected for Galactic 
extinction using the maps by \citet{sch98} and for the effects of seeing and sky level as
in \citet{coenda09}. Absolute magnitudes have been 
computed assuming a flat cosmological model with parameters $\Omega_0=0.3$, 
$\Omega_{\Lambda}=0.7$ and $H_0=100~h~{\rm km~s^{-1}~Mpc^{-1}}$,
$K-$corrected and band-shifted to a redshift $z=0.1$ using the method of 
\citet{blanton03}~({\small KCORRECT} version 4.1). All magnitudes are in the AB system.

\section{The RS and the environment}
%%%%%%%%%%%%%%%%%%%%%%%%%%%%%%%%%%%%%%%%%%%%%%%%%%%%%%%%%%%%%%%%%%%%%%%%%%%%%%%
%%%%%%%%%%%%%%%%%%%%%%%%%%%%%%%%%%%%%%%%%%%%%%%%%%%%%%%%%%%%%%%%%%%%%%%%%%%%%%%
\begin{figure}
\includegraphics[width=90mm]{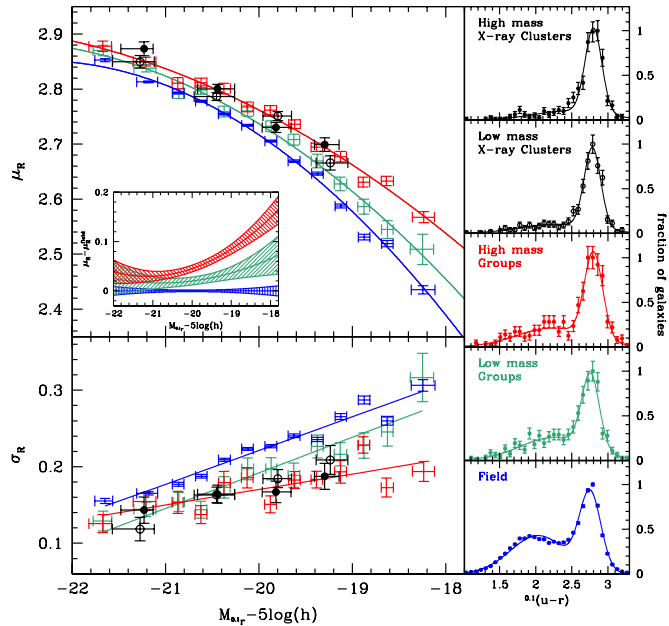}
\caption{Mean colour ($\mu_R$) and scatter ($\sigma_R$) of the RS as a function of absolute magnitude. 
Upper left panel: $\mu_R$ for galaxies in the field (blue), low mass groups (green),
high mass groups (red), low mass clusters (open black circles) and high mass
clusters (filled black circles). The abscissas are the median and the horizontal 
error-bars are the 25\% and 75\%  quartiles of the absolute magnitude distribution within each 
luminosity bin. Vertical error-bars are the 1$\sigma$ error estimates from the fitting procedure. 
The three continuous functions shown are the best fit quadratic models
for field and group galaxies. Inset panel shows the quadratic fits after subtracting
the field's and in shaded areas the 95\% confidence levels. 
Lower left panel: Similar to the upper left panel but showing $\sigma_R$ as a
function of absolute magnitude. 
Continuous functions are the best linear fits for field and group galaxies.
In the small panels at the right, we show
the colour distributions and the two Gaussian fit for the luminosity bins that
include galaxies with $M_{^{0.1}r}-5\log(h)=-20.$}
\label{fig4}
\end{figure}
%%%%%%%%%%%%%%%%%%%%%%%%%%%%%%%%%%%%%%%%%%%%%%%%%%%%%%%%%%%%%%%%%%%%%%%%%%%%%%%
%%%%%%%%%%%%%%%%%%%%%%%%%%%%%%%%%%%%%%%%%%%%%%%%%%%%%%%%%%%%%%%%%%%%%%%%%%%%%%%
\begin{figure}
\includegraphics[width=90mm]{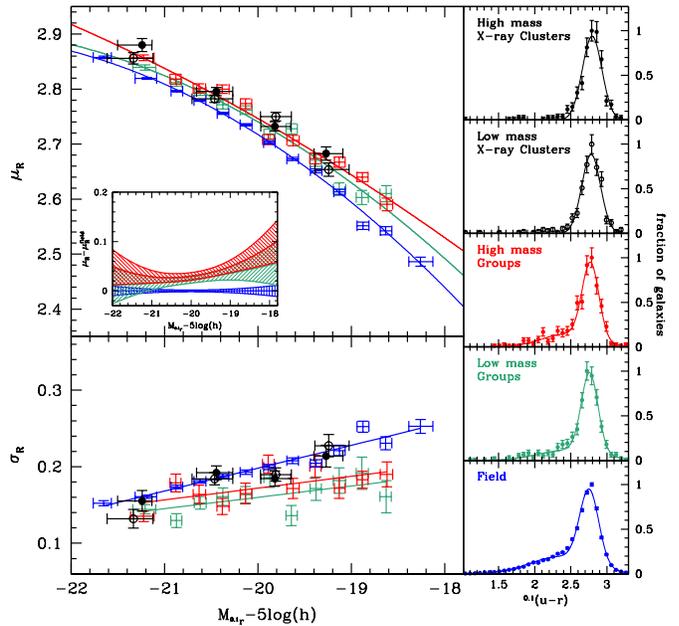}
\caption{Similar to Figure \ref{fig4}, but considering only early-type galaxies
according to their concentration parameter: $C>2.5$.}
\label{fig5}
\end{figure}
%%%%%%%%%%%%%%%%%%%%%%%%%%%%%%%%%%%%%%%%%%%%%%%%%%%%%%%%%%%%%%%%%%%%%%%%%%%%%%%
%%%%%%%%%%%%%%%%%%%%%%%%%%%%%%%%%%%%%%%%%%%%%%%%%%%%%%%%%%%%%%%%%%%%%%%%%%%%%%%

At fixed luminosity, the colour distribution of galaxies is well described by the
sum of two Gaussian functions (e.g. \citealt{baldry04,balogh04,moml06}), 
representing the blue and red galaxy populations. Thus, by fitting this model
to the colour distribution of galaxies at different absolute magnitude 
bins, one can have a suitable quantification of the colour-magnitude
distribution of galaxies as demonstrated by \citet{baldry04}.
In this section, we study the environmental dependence of the RS by fitting the 
two-Gaussian model to the colour distribution of field, group, and cluster galaxies
as a function of absolute magnitude, using a standard Levenberg-Marquardt method.

When analysing the RS of galaxies in our group sample special care must be taken.
Since the MGS is essentially a flux limited sample of galaxies in the range
$14.5\le r \le 17.77$, to maximise the number of galaxies in our analyses, 
we could be tempted to weight each galaxy in order to compensate 
for the Malmquist bias, for instance using the usual $1/V_{\rm max}$ method \citep{schmidt68}. 
This approach, however, has an important bias: fainter galaxies are effectively observed in 
lower redshift groups, i.e., within smaller volumes, and thus, located preferentially
in lower mass systems for all plausible group mass functions.
To avoid this bias, one has to deal with volume limited samples of galaxies.
This is what we do in this work, but not in a straightforward way as would be 
using galaxies in groups within a single volume.
That approach has the inconvenience of, the fainter the faintest galaxies
included in the sample are, the smaller the volume is, and thus, the smaller the 
number of bright galaxies included.
Our approach consists in defining, for each luminosity bin analysed, a volume limited sample of galaxies
belonging to a subset of groups selected in order to have a particular mass distribution.  
The apparent magnitude cut-offs define, for each luminosity bin, a minimum and
a maximum redshift within which the galaxy sample 
is volume limited. We show this in Figure \ref{fig1}.
It is clear that, different luminosity bins sample different volumes. Therefore
the mass distributions of groups will differ from one another. The larger volumes
will include, on average, groups more massive than the smaller volumes.
To have roughly the same group mass distribution from bin to bin, we proceed
as follows. First, we choose a reference bin and compute the mass distribution
of groups in it. Then, for each other bin, we apply a Monte-Carlo algorithm to select 
the largest subset of groups that closely follows the mass distribution on the reference bin.
We use as a reference the luminosity bin $-19.0\le M_{^{0.1}r}-5\log(h)
\le -18.75$ that spans the redshift range $0.02\le z\le 0.067$ (bin III in Figure \ref{fig1}). 
In Figure \ref{fig2} we show in continuous red line the mass distribution of groups in the redshift range 
of the reference bin and the resulting group mass distributions for the remaining 12 bins of Figure \ref{fig1}.

Since our sample of X-ray clusters has only 42 systems after applying the $z=0.12$
cut-off, the procedure above can not be applied. In this case, we use only 4 luminosity bins,
the mass distributions of the parent clusters are shown in Figure \ref{fig3}. A Kolmogorov-Smirnov
test among the mass distributions in Figure \ref{fig3} taken by pairs gives significance levels for 
the null hypothesis that they are drawn from the same distribution above 98\% in all cases.

%%%%%%%%%%%%%%%%%%%%%%%%%%%%%%%%%%%%%%%%%%%%%%%%%%%%%%%%%%%%%%%%%%%%%%%%%%%%%%%
%%%%%%%%%%%%%%%%%%%%%%%%%%%%%%%%%%%%%%%%%%%%%%%%%%%%%%%%%%%%%%%%%%%%%%%%%%%%%%%
\begin{figure}
\includegraphics[width=90mm]{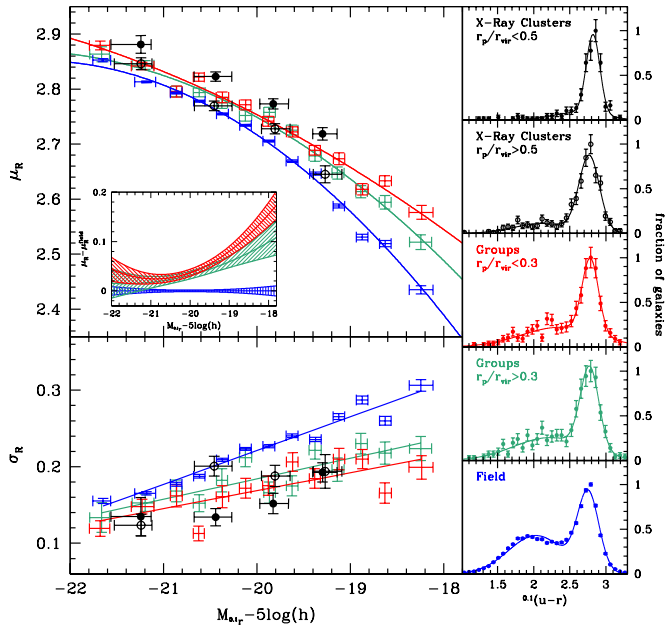}
\caption{Mean colour ($\mu_R$) and scatter ($\sigma_R$) of the RS as a function of absolute magnitude. 
Upper left panel: $\mu_R$ for galaxies in the field (blue), the outer parts of groups (green),
inner parts of groups (red), outer parts of clusters (open black circles) and the inner parts
of clusters (filled black circles). The abscissas are the median and the horizontal 
error-bars are the 25\% and 75\%  quartiles of the absolute magnitude distribution within each 
luminosity bin. Vertical error-bars are the 1$\sigma$ error estimates from the fitting procedure. 
The three continuous functions shown are the best fit quadratic models
for field and group galaxies. Inset panel shows the quadratic fits after subtracting
the field's and in shaded areas the 95\% confidence levels. 
Lower left panel: Similar to the upper left panel but showing $\sigma_R$ as a
function of absolute magnitude. 
Continuous functions are the best linear fits for field and group galaxies.
In the small panels at the right, we show
the colour distributions and the two Gaussian fit for the luminosity bins that
include galaxies with $M_{^{0.1}r}-5\log(h)=-20.$}
\label{fig6}
\end{figure}
%%%%%%%%%%%%%%%%%%%%%%%%%%%%%%%%%%%%%%%%%%%%%%%%%%%%%%%%%%%%%%%%%%%%%%%%%%%%%%%
%%%%%%%%%%%%%%%%%%%%%%%%%%%%%%%%%%%%%%%%%%%%%%%%%%%%%%%%%%%%%%%%%%%%%%%%%%%%%%%
\begin{figure}
\includegraphics[width=90mm]{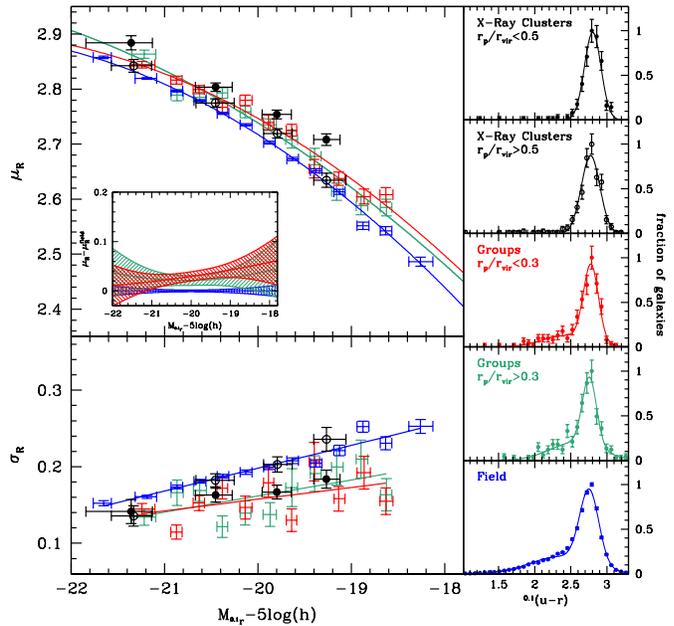}
\caption{
Similar to Figure \ref{fig6}, but considering only early-type galaxies
according to their concentration parameter: $C>2.5$.}
\label{fig7}
\end{figure}
%%%%%%%%%%%%%%%%%%%%%%%%%%%%%%%%%%%%%%%%%%%%%%%%%%%%%%%%%%%%%%%%%%%%%%%%%%%%%%%
%%%%%%%%%%%%%%%%%%%%%%%%%%%%%%%%%%%%%%%%%%%%%%%%%%%%%%%%%%%%%%%%%%%%%%%%%%%%%%%
%%%%%%%%%%%%%%%%%%%%%%%%%%%%%%%%%%%%%%%%%%%%%%%%%%%%%%%%%%%%%%%%%%%%%%%%%%%%%%%
%%%%%%%%%%%%%%%%%%%%%%%%%%%%%%%%%%%%%%%%%%%%%%%%%%%%%%%%%%%%%%%%%%%%%%%%%%%%%%%
\subsection{The mass dependence of the RS}
In Figure \ref{fig4} we show the mean $^{0.1}(u-r)$ colour ($\mu_R$), and the width ($\sigma_R$) 
of the Gaussian function that fits the RS a a function of absolute $^{0.1}r$ magnitude
for field galaxies, and galaxies in groups and clusters of different masses. Error-bars are
1$\sigma$ estimates as given by the fitting procedure.
We split groups and clusters into low and high mass according to the median
of their mass distribution, that is ${\cal M}_{\rm med}^{\rm gr}=1.3\times10^{13} M_{\odot} h^{-1}$
for groups, and ${\cal M}_{\rm med}^{\rm cl}=5\times10^{14} M_{\odot} h^{-1}$, for the 
sample of X-ray clusters.
Field galaxies in each luminosity bin were selected to be volume limited.
In all cases, the abscissas are the medians of the corresponding distribution of absolute magnitudes
in the bin, and the horizontal error-bars are the 25\% and 75\% quartiles.
In the right panels we show  examples of the ability of the method to obtain
good fits to the colour distributions. 

Although the RS has been historically fitted as a linear relation, in more recent years
(e.g. \citealt{baldry04,janz09,skelton09}) it has been found that the relation is best described by  
an $S-$shape.  Modelling the galaxy colour-magnitude diagram, \citet{baldry04} found that 
both $\mu_R$ and $\sigma_R$ follow {\em tanh}-laws as a function of absolute magnitude.  
As can be seen in Figure \ref{fig4}, $\mu_R$ of field and group galaxies follow smooth curves as a 
function of absolute magnitude. We checked that {\em tanh}-laws can provide visually suitable fits to the data points
but, since we do not probe fainter magnitudes where the inflection point of the {\em tanh}-law
is expected, the estimated errors in the resulting parameters were too large and gave rise to
broad confidence bands around the fit. We found that quadratic polynomials provide good fits to our data
points. In Figure \ref{fig4} we show with continuous line the best quadratic fit to the $\mu_R$ for
field, and group galaxies. In an inset panel we show the $95\%$ confidence level of the quadratic fits. 
We did not fit $\mu_R$ of X-ray clusters since we have a much smaller number of galaxies and a cut-off at
$M_{^{0.1}r}-5\log(h)=-19$, and to obtain reliable fits to the colour distributions 
we had to split cluster galaxies into 4 luminosity bins only.

It is clear, from Figure \ref{fig4}, that $\mu_R$ of field galaxies is clearly 
bluer than its group and cluster counterparts. Among groups, $\mu_R$ is systematically
redder for the high mass subsample for galaxies fainter than $M_{^{0.1}r}-5\log(h)\sim -20.5$.
Regarding the scatter of the RS, $\sigma_R$, it shows a tendency of becoming broader with decreasing luminosity
in all environments probed (lower left panel of Figure \ref{fig4}). 
We also show linear fits to $\sigma_R$ as a function of $M_{^{0.1}r}$ for field
and group galaxies. Field galaxies show the highest 
values of $\sigma_R$ for the whole range of absolute magnitudes.  Among galaxies in groups,
$\sigma_R$ is smaller for the high group mass subsample, with the exception of one single bin.
The points corresponding to galaxies in clusters are consistent with the values from high mass groups.

Figure \ref{fig5} is similar to Figure \ref{fig4} but considers red early type galaxies alone. 
To do this we repeated the fitting procedure only
to galaxies that have concentration parameters $C>2.5$ \citep{st01} in the $r-$band. 
The exclusion of these galaxies did not completely remove the blue population 
as can be seen in the right panels of Figure \ref{fig5}.
For groups, the differences observed between low and high mass subsamples is not clear anymore.
We were not able to keep the same number
of bins as in Figure \ref{fig4} since the exclusion of an important number of galaxies 
limited the ability of the method for obtaining reliable Gaussian fits.
For X-ray clusters, the results were not affected by the $C$ cut-off, since only
a few galaxies in the sample have $C<2.5$.
What is still clear from Figure \ref{fig5} is that field galaxies have $\mu_R$
values that are systematically bluer that their counterparts in systems of galaxies.
Regarding $\sigma_R$, field values are consistent with cluster's and there is no clear 
distinction between groups of different masses.

\subsection{The system-centric dependence of the RS}
We also explore the dependence of the RS with the distance to the 
centres of the systems in which the galaxies are located. 
We use the same samples of galaxies constructed for the analyses of the previous subsection.
We split galaxies in systems according to whether their projected
distance to the system's centre in units of the virial 
radius, $r_p/r_{\rm vir}$, is larger or smaller than the sample's median. 
The median values are 0.3 for groups, and 0.5 for X-ray clusters. 

In Figure \ref{fig6} we show $\mu_R$ and $\sigma_R$ as a function of absolute magnitude
for galaxies in the inner and outer regions of systems, classified by their
$r_p/r_{\rm vir}$ ratios. We also show as a comparison, the field values 
from Figure \ref{fig4}. For group galaxies, $\mu_R$ in the inner regions
is systematically redder than its outer regions counterpart for $M_{^{0.1}r}-5\log(h)\ge -19$.
Even with larger error-bars, $\mu_R$ in the inner regions of clusters 
is significantly redder than in the outer regions for all luminosities probed.
It is also clear that $\mu_R$ in the field is distinctively bluer than
in groups and in the inner regions of clusters for  $M_{^{0.1}r}-5\log(h)\sim -20.5$.
It is interesting to notice that the $\mu_R$ in the outer regions of clusters
is consistent with that of field galaxies.
The scatter of the RS is systematically larger for field galaxies over the whole range of magnitudes.
Despite noisier trends, $\sigma_R$ is marginally larger in the outer parts of groups 
than in the inner regions. For galaxies in clusters larger $\sigma_R$ values are found in the outer regions.

Figure \ref{fig7} shows the resulting sequences for red early type galaxies only. 
The differences in the RS between the inner and the outer parts of groups disappear.
As in the previous subsection, since a vast majority of our cluster
galaxies are early type, there are no significant changes between the
cluster sequences shown in Figures \ref{fig6} and \ref{fig7}.
Lower left panel of Figure \ref{fig7} shows that the scatter of the RS is larger in the
field and in the outskirts of clusters and no clear distinction can be made between
inner and outer parts of groups.
%%%%%%%%%%%%%%%%%%%%%%%%%%%%%%%%%%%%%%%%%%%%%%%%%%%%%%%%%%%%%%%%%%%%
%%%%%%%%%%%%%%%%%%%%%%%%%%%%%%%%%%%%%%%%%%%%%%%%%%%%%%%%%%%%%%%%%%%%
\section{Discussion and conclusions}
Using volume limited samples of bright field, group and X-ray cluster galaxies we study 
the environmental dependence of the red galaxy sequence. Field 
and system galaxies were drawn from the DR7 of the SDSS. Properties of the colour-magnitude 
relation of red galaxies are studied by fitting a two-Gaussian model to the colour distribution 
as a function of the absolute magnitude. Based on the concentration parameter, we have also 
selected red early type galaxies. The resulting RSs have curved shapes that are well described 
by quadratic polynomials over the luminosity range probed.

We found that for each luminosity, mean colours of field red-galaxies are bluer than their
group and cluster counterparts. Among systems of galaxies, high mass groups and 
clusters show the reddest RS. Red galaxies in low mass groups have intermediate colours 
between field and cluster galaxies. The differences in the mean colours are larger for 
the faintest luminosity bins. If red early-type galaxies are considered, galaxies in groups
and clusters populate the same RS. Nevertheless, for the 
whole range of magnitudes, field galaxies are bluer than galaxies in systems.
It is important to note that the RS defined by early type galaxies has properties
different than those obtained for red galaxies without morphological selection, suggesting 
different evolutionary paths for non star forming galaxies of different morphological types.
We have also analysed the scatter around the mean RS. $\sigma_R$ presents, for all 
environments analysed in this work, a systematic increase towards fainter magnitudes. 
Galaxies in groups and clusters show a similar behaviour, while field galaxies present for the whole range of 
magnitudes the highest values of the intrinsic scatter. This property of field galaxies is consistent 
with a more complex history of the star formation and quenching processes. 

If the system-centric distance is used to characterise the environment, 
our results are slightly different. With the exception of a few luminosity bins, galaxies in the 
core and the outskirts of groups show similar RSs whereas galaxies in X-ray clusters show 
a strong dependence on $r_p/r_{vir}$ for the whole range of magnitude, being inner galaxies 
the reddest objects.  Similar results are found for red early-type galaxies. It should be 
noted that galaxies in the outskirts of clusters have colours similar to those of field 
galaxies. This is in agreement with the scenario of field galaxies falling into clusters.
If the scatter is analysed, the differences between system mass and system-centric distance 
are more significant. Field galaxies are still the objects that show the highest values of scatter,
nevertheless, red early type galaxies in the outskirts of X-ray clusters show a scatter as large as
field galaxies', while galaxies in the core of clusters have lower values that are
comparable to those of group galaxies.

Our results indicate that the RS depends on the present time environment. Nevertheless, 
most of the stars that place a galaxy in the RS were formed when the universe was very 
young, moreover, most of the physical processes acting at $z\sim 0$ were not necessarily present 
at times when the bulk of the stellar populations were forming. 
Therefore, the origin of the dependence of 
the RS on the present time environment should be looked for at high redshift. 
\citet{steidel05} found that galaxies in protoclusters have mean ages $\sim 2$ times 
larger than identical galaxies outside the protocluster structures. This observational 
evidence is in agreement with the $\Lambda$CDM model where denser dark matter 
concentrations collapse earlier and form an important fraction of their stellar mass in  
early short events \citep{springel05}. These differences in age 
will imply subtle differences in the present day colours. In order to test this hypothesis, 
we have used the code {\small GALAXEV} by \citet{bc03} to construct the SED of 
a solar metallicity Salpeter IMF stellar population with a star formation rate of the form 
$\psi(t)\propto \exp(-t/\tau)$, and assumed $\tau=1$Gyr. This model SED allows us to relate 
colour differences in the RS of different environments with age differences, assuming that 
all galaxies in the RS are represented by such a simple model.  
Within this scenario, the colour differences between field and high mass group galaxies
in Figure \ref{fig1} imply age differences ranging from $200$ to $600$Myr in the magnitude 
range $-22\leq M_{^{0.1}r}-5\log(h)\leq-18$. 
These differences are comparable to those found by \citet{delucia06} when analysing
the formation history of elliptical galaxies in halos of different masses. 

We conclude that, the RS shows dependencies with environment that could be mostly 
interpreted in terms of the typical age of the bulk of their stellar population. 
Beyond the fact that each environment
shows a different RS, within a given environment, the RS scatter also shows variations. 
If these results are also analysed in terms of ages, assuming that higher 
scatter corresponds to younger ages (Kodama and Arimoto 1997), our results indicate that the oldest 
stellar population are to be found in the brightest galaxies. 
Field galaxies and cluster members in the outskirts have values of scatter 
equal or larger than the corresponding values for groups and galaxies in the core of clusters, 
also consistent with the scenario of field galaxies falling into clusters. The same effect seems to be 
present in the outskirts of groups of galaxies. 
However this can only be as a hint, given the significant dispersion observed, particularly in the mean 
values of $\sigma_R$. 
If red early type galaxies are considered, the differences in the RSs are smaller, moreover, 
we found that bright galaxies ($M_{^{0.1}r} <-21$) have the same colour (ages) irrespective of the environment. 
In other words, bright red early type galaxies are the only 
objects that reached nearly the same evolutionary stage in groups, clusters and in the field.
The clear difference between RS galaxies in the core and the outskirts of clusters, suggests that 
the cluster environment could be playing a roll in the drift of galaxies of intermediate 
magnitudes towards redder colours, producing an increase of the fraction of red galaxies. 
Nevertheless, the fact that galaxies in groups and in the core of X-ray clusters populate nearly 
the same RS (the cluster RS is slightly redder), suggests that the cluster environment is not 
necessary to populate the RS. Our findings give support to a scenario in which the RS of galaxies in clusters 
is made of two population of galaxies: on one hand, very old objects that are those observed galaxies populating
the RS in clusters at $z\sim1$ and that were formed in proto-clusters, and, on the other hand, 
galaxies that evolved in groups or in the field with a more complex history of mergers and star 
formation that ended up in clusters at lower redshift. 
These two populations that can not be distinguished by their colour alone, would show
differences in their morphology.
 
%%%%%%%%%%%%%%%%%%%%%%%%%%%%%%%%%%%%%%%%%%%%%%%%%%%%%%%%%%%%%%%%%%%%%%%%%%%%%%%
%%%%%%%%%%%%%%%%%%%%%%%%%%%%%%%%%%%%%%%%%%%%%%%%%%%%%%%%%%%%%%%%%%%%%%%%%%%%%%%
\section*{Acknowledgements}
We thank the anonymous referee for comments and suggestions that improved
this paper.
We are indebted to Ariel Zandivarez for the group identification in
the DR7 MGS.
HJM acknowledges the support of a Young Researcher's grant from Agencia 
Nacional de Promoci\'on Cient\'\i fica y Tecnol\'ogica Argentina, PICT 
2005/38087. This work has been partially supported with grants from Consejo 
Nacional de Investigaciones Cient\'\i ficas y T\'ecnicas de la Rep\'ublica 
Argentina (CONICET) and Secretar\'\i a de Ciencia y Tecnolog\'\i a de la 
Universidad de C\'ordoba.

Funding for the Sloan Digital Sky Survey (SDSS) has been provided by the 
Alfred P. Sloan Foundation, the Participating Institutions, the National 
Aeronautics and Space Administration, the National Science Foundation, the U.S.
Department of Energy, the Japanese Monbukagakusho, and the Max Planck Society. 
The SDSS Web site is http://www.sdss.org/. The SDSS is managed by the 
Astrophysical Research Consortium (ARC) for the Participating Institutions. 
The Participating Institutions are The University of Chicago, Fermilab, the 
Institute for Advanced Study, the Japan Participation Group, The Johns Hopkins
University, the Korean Scientist Group, Los Alamos National Laboratory, the 
Max Planck Institut f\"ur Astronomie (MPIA), the Max Planck Institut f\"ur 
Astrophysik (MPA), New Mexico State University, University of Pittsburgh, 
University of Portsmouth, Princeton University, the United States Naval 
Observatory, and the University of Washington.
%%%%%%%%%%%%%%%%%%%%%%%%%%%%%%%%%%%%%%%%%%%%%%%%%%%%%%%%%%%%%%%%%%%%%%%%%%%%%%%
%%%%%%%%%%%%%%%%%%%%%%%%%%%%%%%%%%%%%%%%%%%%%%%%%%%%%%%%%%%%%%%%%%%%%%%%%%%%%%%

\label{lastpage}
\end{document}